\documentclass{aastex}
\usepackage{spr-astr-addons}

\usepackage{graphics}
\usepackage{epsfig}

\begin{document}

\title{Stellar dynamics in young clusters: the formation of massive runaways and very massive runaway mergers}

%% Running heads

\author{D. Vanbeveren\altaffilmark{1,2}, H.
Belkus\altaffilmark{1}, J. Van Bever\altaffilmark{3}, N. Mennekens\altaffilmark{1}}

\altaffiltext{1}{Astrophysical Institute, Vrije Universiteit Brussel, Brussels, Belgium, \\dvbevere@vub.ac.be,
hbelkus@vub.ac.be, nmenneke@vub.ac.be}
\altaffiltext{2}{Groep T, Association K.U.Leuven, Leuven, Belgium}
\altaffiltext{3}{Institute of Computational Astrophysics, St.-Mary's University, Halifax, Canada,
\\vanbever@penguin.stmarys.ca}

\begin{abstract}
In the present paper we combine an N-body code that simulates the dynamics of
young dense stellar systems with a massive star evolution handler that
accounts in a realistic way for the effects of stellar wind mass loss. We discuss two topics: 

\noindent 1. The formation and the evolution of very massive stars (with a mass
$>$ 120 $M_\odot$) is followed in detail. These very massive stars are formed in the cluster core as a
consequence of the successive (physical) collison of 10-20 most massive stars of the cluster (the process is
known as runaway merging). The further evolution is governed by stellar wind mass loss during core hydrogen
burning and during core helium burning (the WR phase of very massive stars). Our simulations reveal that
as a consequence of runaway merging in clusters with solar and supersolar
values, massive black holes can be formed but with a maximum mass $\approx$ 70 $M_\odot$. In
small metallicity clusters however, it cannot be excluded that the runaway
merging process is responsible for pair instability supernovae or for the
formation of intermediate mass black holes with a mass of several 100 $M_\odot$.

\noindent 2. Massive runaways can be formed via the supernova explosion of one of the components in a binary
(the Blaauw scenario) or via dynamical interaction of a single star and a binary or between two binaries in a
star cluster.  We explore the possibility that the most massive runaways (e.g., $\zeta$ Pup, $\lambda$ Cep,
BD+43$^{o}$3654) are the product of the collision and merger of 2 or 3 massive stars. 
\end{abstract}


\section{Introduction}

The temporal evolution of the population of different types of massive stars in starbursts has been 
the subject of numerous research topics in the last two decades. We can distinguish studies dealing with
OB-type stars, Wolf-Rayet (WR) type stars and compact objects. Some of them have only been concerned with
single stars (e.g. Arnault et al. 1989; Maeder 1991; Mas-Hesse \& Kunth 1991; Cervino \& Mas-Hesse 1994;
Meynet 1995; Leitherer et al. 1999). The effects of close binaries were investigated by Dalton \& Sarazin
(1995); Schaerer \& Vacca (1998); Pols et al. (1991); Van Bever \& Vanbeveren (1997, 2000, 2003); Vanbeveren
et al. (1997, 1998a, b, c); Belczynski et al. (2002). However, none of the papers listed above account for
the effects of stellar dynamics, although young massive starbursts may be very dense. Attempts to include
dynamics in order to follow the early evolution of massive starbursts have been presented by Portegies Zwart
et al. (1999), Ebisuzaki et al. (2001); Portegies Zwart \& McMillan (2002); G\"urkan et al. (2004); Freitag
et al. (2006). These papers investigate the starburst conditions to initiate a runaway collision (runaway
merger), which may lead to the formation of a very massive object at the centre of the cluster, and the
possible formation of an intermediate mass black hole (IMBH). IMBHs and their formation became a hot topic
soon after the ROSAT and Einstein X-ray surveys of galaxies, when it was realised that the high-luminosity
sources (with luminosity up to
$\sim$10$^{42}$ erg/s) were linked to regions of intense star formation activity, starbursts. Despite the
limited sensitivity and spatial resolution of both telescopes, the images of nearby galaxies suggested the
presence of highly super-Eddington stellar-mass sources with luminosities as high as $\sim$10$^{40}$
erg/s. The X-ray telescope on board of Chandra confirmed the existence of these sources, but it also revealed
the existence of individual sources with a luminosity $\sim$10$^{41}$-10$^{42}$ erg/s (Ptak \&
Colbert, 2004 for a review of galaxies with Ultra Luminous X-ray sources, ULXs). Many models to explain ULXs
have been proposed in literature (e.g., Fabbiano, 1989; Colbert \& Mushotsky, 1999; Perna \& Stella, 2004, and
references therein) and IMBHs is one of them. MGG 11 is a young dense star cluster $\sim$200 pc from the
centre of the starburst galaxy M 82, whose parameters have been studied by McCrady et al. (2003). A ULX
associated with the cluster, the runaway collision process and formation of an IMBH in order to explain this
source was promoted by Portegies Zwart et al. (2004). The question whether or not an IMBH is needed in order
to explain ULXs was addressed in detail by Soria (2007), who concluded that most of them can be explained
with a 50-100 $M_\odot$ BH accreting mass at super-Eddington rates. 

At the center of the  Galactic bulge lies a supermassive black hole (SMBH with a mass
$\sim$3-4.10$^{6}$ $M_\odot$, Ghez et al., 1998, 2000). Several young ($\le$ 10 Myr)
dense star clusters were observed within $\sim$100 pc of this SMBH (Arches, Figer et al., 2002; Quintuplet,
Figer et al., 1999; IRS 13E, Maillard et al., 2004; IRS 16SW, Lu et al., 2005). The formation of IMBHÕs in
bulge clusters of this type has been investigated by Portegies Zwart et al. (2006). The authors concluded
that IMBHÕs that are formed as a consequence of core collapse accompanied by runaway star collisions in dense
clusters with the properties of bulge clusters, may be the building blocks of SMBHÕs, a model that was
originally proposed by Ebisuzaki et al. (2001). 

The runaway merger process in dense stellar systems as a model to explain IMBHs
contains two major uncertainties: first, the core collapse and the formation of a runaway merger can be
considered as facts, but it is as yet unclear whether or not a very massive object like this will ever become
a star, and, secondly, when this object becomes a very massive star, what the effect of stellar wind mass
loss is on its evolution and on the final mass the moment that the star collapses. The evolution of very
massive stars has been discussed recently by Belkus et al. (2007) and it was concluded that stellar wind mass
losses during core hydrogen burning and during core helium burning are very important. A convenient
evolutionary recipe for such very massive stars was presented, a recipe that can easily be implemented in a
N-body dynamical code. In the present paper we first investigate the formation and evolution of
very massive stars in a dense cluster environment. In section 2 we present our N-body code, whereas
section 3 summarizes the massive star evolution handler. In section 4 we present simulations of the
dynamical evolution of a cluster combined with the massive and very massive evolution handler, and we discuss
the results in relation to the cluster MGG 11.

Massive star runaways are defined as massive stars with a peculiar space velocity $\ge$ 30-40 km/s. At least
10\% of the O-type stars are classified as runaways (Gies \& Bolton, 1986). They can form either by the
dynamical ejection from a cluster due to single star-binary or binary-binary interactions (Poveda et
al., 1967; Leonard \& Duncan, 1988, 1990), or by the explosion as a supernova (SN) of a member of a close
binary (Blaauw, 1961). A word of caution: it is not because a runaway is observed close to a dense cluster
that one should favor the dynamical formation mechanism. If most of the stars are formed in clusters, also
binary-SN  formed runaways will come out of a cluster. On the other hand, it is not because there is no
O-type cluster observed in the neighbourhood of an O-type runaway that one should favor the binary-SN
mechanism. Dynamically formed runaways are in many cases rejuvenated collision products (stellar mergers of
2-3 stars, see section 5) and the other O-type stars of the parent cluster of the runaway may have
disappeared already.

$\zeta$ Pup, $\lambda$ Cep and BD+43$^{o}$3654 are 3 most massive runaways with a runaway velocity between
40 km/s and 70 km/s (Vanbeveren et al., 1998b, c; Hoogerwerf et al., 2001; Comeron \& Pasquali, 2007). In
section 5 we explore the dynamical ejection process in order to explain their properties.   

\section{The dynamical N-body code for young dense stellar systems}

A full description of our N-body code will be presented elsewhere 
(Van Bever et al., 2008). Summarizing, our code (written in C++) contains
three main components. There is the standard scheme for integrating single
stars and the centers of mass of stellar hierarchies (e.g., binaries), for
which we use the fourth order Hermite scheme described by Makino \& Aarseth (1992). The
treatment of compact subsystems requires special techniques that deal with
the extremely small inter-particle distances that occur in these cases. We
distinguish between two-body motion, which is handled by the
Kustaanheimo-Stiefel regularisation technique, and more complex encounters
between more than 2 stars. The latter is treated by the so-called chain
regularization (Mikkola \& Aarseth, 1993, and references therein), which
allows the accurate integration of a compact subsystem with arbitrary
membership. 
 
The code generates a zero age massive single star population according to a 
predefined IMF and a predefined spatial cluster distribution. We use a King
model with various concentrations, parametrized by the dimensionless central
potential, W$_{0}$. We do not account for the presence of a primordial binary
population in the cluster simulations of the present paper. 

\begin{figure}
			\begin{center}
				\epsfig{file=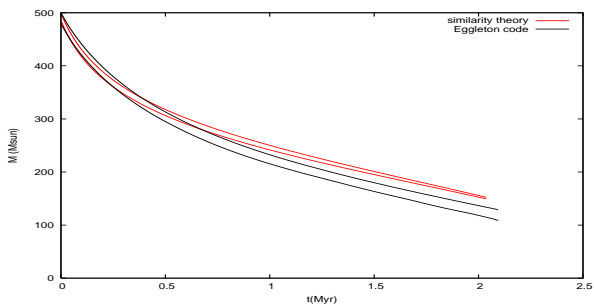, height=4cm,width=8cm}
				\end{center}
				\caption[]{The temporal evolution during core hydrogen burning of the total mass and of the core mass of a
				500 $M_\odot$ star calculated with the Eggleton stellar evolutionary code (black color) and with the very
				massive star evolutionary recipe discussed by Belkus et al. (2007) (red color).}
\end{figure}

\section{The massive single star evolution handler}

Our preferred evolutionary model for massive single stars has been described
in a number of papers (Vanbeveren et al. 1998a, b, c; Van Bever \&
Vanbeveren, 1997, 2000, 2003 and references therein). It is summarized
here together with a few updates.

\begin{enumerate}
\item Stars with an initial mass larger than 40 $M_\odot$ evolve according to the 
LBV scenario as it was introduced by Vanbeveren (1991). Summarizing: due to
the fact that no yellow or red supergiants (YSG and RSG) are observed
brighter than M$_{bol}$ = -9.5, we use as a working hypothesis in our
evolutionary computations that {\it the $\dot M$ during the LBV phase of a star with M$_{bol}$
 $\leq$ -9.5 must be sufficiently large to suppress a large expansion, hence to
prohibit the redward evolution in the HR-diagram}. When this criterion is
implemented into a stellar evolutionary code, the code calculates the mass
loss rates that are needed at any time in order to prohibit the redward
evolution.  Since we do not observe RSGÕs with M$_{bol}$ $\leq$ -9.5 in the LMC or SMC
either, we consider this as evidence that the LBV scenario is independent
from the initial metallicity.

\item The RSG evolutionary phase of massive single stars with an initial
mass $<$ 40 $M_\odot$ is computed in most of the present single star evolutionary codes by using the de Jager
et al. (1988) stellar wind mass loss rate formalism (e.g., Meynet \& Maeder, 2003; Eldridge \& Vink, 2006).
Vanbeveren (1995) and Vanbeveren et al. (2007) illustrated that an update may be necessary, that affects
the evolution of single stars with an initial mass between 20 $M_\odot$ and 40 $M_\odot$. Our evolutionary
handler accounts for this update.

\item Since 1998 we use core helium burning WR mass loss rates in our
evolutionary code, which correspond to empirical rates determined by
accounting for the effects of clumping. The WR rates are assumed to be
Z-dependent (Vanbeveren et al., 1998b, c; Van Bever \& Vanbeveren, 2003). Notice that the WR-mass loss
rate formalism critically affects the pre-supernova mass of a massive star,
thus also the final fate (neutron star or black hole) and, in case of a black
hole, the mass of this compact object. 

\item Realistic dynamical simulations of young dense systems of massive
stars reveal the existence of what can be considered as one of the most
spectacular events in astrophysics: the gravitational encounter of two
objects (single star-single star), (single star-binary) or (binary-binary)
resulting in many cases in a physical collision of two stars. A collision of
two massive stars in dense stellar environments may initiate a chain
reaction where the same collision object merges with other massive stars: the
term Ôrunaway mergerÕ is used. In many cases, this runaway merger may become
as massive as 1000 $M_\odot$ or even larger (Portugies Zwart et al., 2006, and
references therein, see also section 4) and to investigate the consequences
of this process it is indispensable to know how such an object further
evolves. Using 3D smoothed-particle-hydrodynamics (SPH) Suzuki et al. (2007)
simulated the collision and merging of 2 massive stars. The evolution during
merging depends on the mass ratio of the two colliding stars, but after the
thermal adjustment (Kelvin-Helmholtz contraction) the merger is nearly
homogenized. In our N-body
simulations we therefore assume that massive collision objects are
homogenized and become ZAMS stars (with the appropriate chemical composition)
on a timescale which is short compared to the stellar evolutionary
timescale. Obviously, the evolution of this new massive star will be
critically affected by stellar wind mass loss.

\medskip

\noindent INTERMEZZO: Suzuki et al. (2007) simulated the collision of a star with a mass = 88 $M_\odot$ with a
star with the same mass and one with a mass = 28 $M_\odot$. The merging process is very short (of the order of
days) and, interestingly, during the merging the collision object loses $\sim$ 10 $M_\odot$. It is tempting to
link this collision and merging process to the $\eta$ Car outburst in the 19th century. Note that a
collision between a single star and a binary may also explain the observed anomalous eccentricity of the
$\eta$ Car binary.    

\end{enumerate}

With our evolutionary library of massive single stars, it is straightforward to estimate the evolution of a
merger with a mass smaller than 120
$M_\odot$, given its chemical composition after homogenization. However, what about
mergers with a mass larger than 120 $M_\odot$, up to 1000 $M_\odot$ (very massive stars =
VMSÕs)? The evolution of VMSs has been studied by Belkus et al. (2007) where
it was demonstrated that stellar wind mass loss plays a crucial role. At
solar metallicity and larger, VMSs lose most of their mass on a timescale of
the order of 2 million years and end their life as a black hole with a mass
less than 75 $M_\odot$. We extended these calculations for stars with an initial mass up to 3000 $M_\odot$.
Stars with a larger mass have a larger luminosity and thus a larger stellar wind mass loss. Our computations
reveal that a 3000 $M_\odot$ star has a mass loss of the order of 10$^{-2}$ $M_\odot$/yr. As a consequence,
the final mass at the end of core helium burning is less than $\sim$ 75 $M_\odot$ as well.  At Z = 0.001 (and
smaller), the final mass may be a factor 2-3 larger and the formation of an IMBH with a mass of a few hundred
$M_\odot$ is a possibility. We presented a convenient evolution recipe for VMSÕs that can
easily be implemented in a dynamical-population code. To illustrate the
validity of our recipe, figure 1 compares evolutionary results of a galactic
500 $M_\odot$ VMS calculated with the Eggleton code with the evolution predicted by
the recipe. As can be noticed, the overall evolutionary results that are important in order to follow the
evolution of this object in a dense cluster are very similar.

\section{The formation and evolution of runaway mergers in dense stellar
systems}

We simulated the early evolution of a dense cluster core containing 3000 
massive single stars (with a mass between 10 $M_\odot$ and 120 $M_\odot$ satisfying the
Salpeter initial mass function) distributed according to a King model (Wo =
9). We adopt a half-mass radius = 0.5 pc. In figure 2 we show the growth in
mass of the collision runaway star with time. Our choice of the half mass
radius was motivated by the fact that, with the stellar wind mass loss
formalism as the one used by Portegies Zwart et al. (2004, 2006), we confirm
the mass evolution of the runaway merger and the possible formation of an IMBH
with a mass as large as 1000 $M_\odot$. However, with a
more realistic mass loss rate formalism, the figure illustrates that if
a dense cluster has solar or supersolar metallicity the formation of an IMBH
is rather unlikely. The BHÕs in our simulation have a mass not larger than 75
$M_\odot$ (see also Belkus et al., 2007). Notice that this may be sufficient in
order to explain the presence of a ULX in MGG 11, provided that we accept the
X-ray formation scenario of Soria (2007).  

In order to investigate the effect of the metallicity on the formation and evolution of runaway mergers
(through the effect of Z on the stellar wind mass loss), we followed the
dynamical evolution of the same cluster as the one discussed above, but for Z
= 0.001. The results are shown in figure 2 as well. The final mass is of the order of 200 $M_\odot$ and we
conclude that the formation of an IMBH is possible in dense clusters with small Z. When small Z globular
clusters are preceded by a massive supercluster phase, it can thus not be excluded that during this
early phase an IMBH formed. 

Notice that pair instability supernovae are expected to happen when the final mass after core helium burning
is between $\sim$ 65 $M_\odot$ and $\sim$ 130 $M_\odot$ (Heger \& Woosley, 2002). The results which are
depicted in figure 2 illustrate that in dense clusters with subsolar metallicity pair instability
supernovae may happen.

More details of the N-body simulations discussed above are given in Belkus et al. (2008).

\begin{figure*}	
				\begin{center}
				\epsfig{file=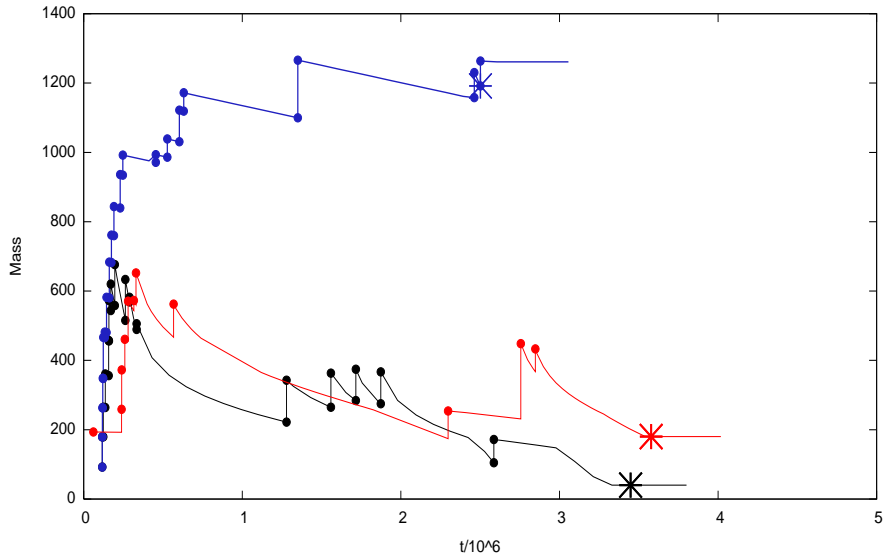, height=8cm,width=12cm}
				\end{center}
				\caption[]{The effect of stellar wind mass loss on the temporal evolution of the runaway merger during
core hydrogen and core helium burning. The blue curve corresponds to a very massive star stellar wind mass
loss rate formalism as the one used by Portegies Zwart et al., the black illustrates the evolution with the
formalism discussed in Belkus et al. (2007) and metallicity Z = 0.02; the red curve is similar as the black
one but for Z = 0.001.  }
   \end{figure*}

\section{The formation of massive runaways}

$\zeta$ Pup, $\lambda$ Cep and BD+43$^{o}$3654 are 3 massive runaways with a runaway velocity between 40
km/s and 70 km/s. Their location in the HR diagram suggests that they belong to the most massive star sample
of the solar neighborhood (Vanbeveren et al., 1998b, c; Hoogerwerf et al., 2001; Comeron
\& Pasquali, 2007). Runaways can be formed by the binary-SN scenario (Blaauw, 1961) where the original
massive primary (the mass loser when the Roche lobe overflow process happens) explodes and eventually
disrupts the binary, leaving a neutron star remnant and a runaway secondary (the mass gainer when the Roche
lobe overflow happens). Such a scenario for $\zeta$ Pup was presented by Vanbeveren et al. (1998b, c). To
explain the significant surface helium enrichment of the star, its rapid rotation and its runaway
velocity (= 70 km/s), the mass transfer phase and the accretion process must be accompanied by spinning-up
and quasi-homogenization of the mass gainer (the full mixing model as it was introduced by Vanbeveren and De
Loore, 1994) whereas the overall evolution should have resulted in a pre-SN binary with a period of the
order of 4 days. The latter requires some fine-tunning. 

To illustrate that the dynamical ejection mechanism is a very valuable alternative, the
FEWBODY software of Fregeau et al. (2004) was used to reproduce the observed properties of $\zeta$ Pup. We
performed over 1 million single star-binary and binary-binary scattering experiments. The details of
these experiments are given elsewhere (Belkus et al., 2008). We explored the effects of different masses and
different binary periods and eccentricities and, obviously, many experiments reproduce $\zeta$ Pup, but to
obtain  a runaway velocity as observed the binaries participating in the scattering process always have to
be very close (periods smaller than 100 days). Most interestingly, in all our experiments, {\it $\zeta$ Pup
turns out to be a merger of 2 or 3 stars.}

\section{Conclusions}

In the present paper we first discuss the dynamical formation (due to runaway merging) and evolution of very
massive stars (with masses up to 1000 $M_\odot$ and more) in the cores of young dense
clusters. To predict whether such a very massive object becomes a stellar mass black hole, an intermediate
mass black hole or explodes as a pair instability supernova, one has to combine a dynamical N-body
code with a massive and very massive star evolutionary library, that considers in detail the importance of
stellar winds and of the metallicity dependence of these winds on the core hydrogen burning and core helium
burning evolution of the massive and very massive stars. Secondly we present arguments in favour of the
dynamical ejection scenario in order to explain the runaways (space velocity larger than 30 km/s) with a mass
larger than 40 $M_\odot$, like $\zeta$ Pup, $\lambda$ Cep and BD+43$^{o}$3654. We conclude:

\begin{enumerate}

\item In clusters with solar or supersolar metallicity, black holes form with a mass smaller than 70-75
$M_\odot$, but the formation of an intermediate mass black hole with a mass of several 100 $M_\odot$ is less
likely.

\item Due to the metallicity dependence of the stellar wind mass loss, the occurence of pair instability
supernovae or the formation of an intermediate mass black hole in dense clusters becomes more probable for
smaller metallicities.

\item It is plausible that at least some of the most massive runaways, like
$\zeta$ Pup, $\lambda$ Cep and BD+43$^{o}$3654, are formed during a dynamical encounter of a massive single
star and a massive close binary, or by two massive close binaries. In this case, the runaway is the merger of
2 or 3 massive stars.
  
\end{enumerate} 

\medskip

\noindent {\bf Ackowledgement}

\noindent We thank Dr. Lev Yungelson who calculated the evolution of the 500 $M_\odot$ star with his
version of the Eggleton code but with our preferred stellar wind mass loss formalism.


\begin{thebibliography}{}

\bibitem{} Arnault, P., Kunth, D., \& Schild, H. 1989, A \& A, 224, 73
\bibitem{} Belczynski, K., Kalogera, V., \& Bulik, T. 2002, Ap. J., 572, 407 

\bibitem{} Belkus, H., Van Bever, J., \& Vanbeveren, D. 2007, Ap. J. 659, 1576
\bibitem{} Belkus, H., Van Bever, J., \& Vanbeveren, D. 2008, in preparation
\bibitem{} Blaauw, A. 1961, Bull. Astron. Inst. Netherlands, 15, 265
\bibitem{} Cervino, M., \& Mas-Hesse, J. M. 1994, A \& A, 284, 749 
\bibitem{} Colbert, E. J. M., \& Mushotzky, R. F. 1999, Ap. J., 519, 89 
\bibitem{} Comeron, F., \& Pasquali, A. 2007, A \& A, 467, L23 
\bibitem{} Dalton, W. W., \& Sarazin, C. L. 1995, Ap.J., 448, 369 
\bibitem{} de Jager, C., Nieuwenhuijzen, H., \& van der Hucht, K. A. 1988, A \& A Suppl. Ser., 72,
259  
\bibitem{} Ebisuzaki, T., et al., 2001, Ap. J., 562, L19 
\bibitem{} Eldridge, J. J., \& Vink, J. S. 2006, A \& A, 452, 295 
\bibitem{} Fabbiano, G. 1989, Annual review of astronomy and astrophysics. Volume 27, 87  
\bibitem{} Figer, D. F., McLean, I. S., \& Morris, M. 1999, Ap. J., 514, 202
\bibitem{} Figer, D.F. et al., 2002, Ap. J., 581, 258 
\bibitem{} Fregeau, J.M. et al., 2004, MNRAS, 352, 1
\bibitem{} Freitag, M., Rasio, F. A., \& Baumgardt, H. 2006, MNRAS, 368, 141 
  
\bibitem{} Ghez, A. M., Klein, B. L., Morris, M., \& Becklin, E. E. 1998, Ap. J., 509, 678 
\bibitem{} Ghez, A. M., et al., 2000, Nature, 407, 349     
\bibitem{} Gies, D. R., \& Bolton, C. T. 1986, Ap. J., 304, 371  
\bibitem{} G\"urkan, M. A., Freitag, M., \& Rasio, F. A. 2004, Ap. J., 604, 632 
\bibitem{} Heger, A., \& Woosley, S. E. 2002, Ap. J., 567, 532 
\bibitem{} Hoogerwerf, R., de Bruijne, J. H. J., \& de Zeeuw, P. T. 2001, A \& A, 365, 49 
\bibitem{} Leitherer, C., et al., 1999, Ap. J. Suppl. Ser., 123, 3 
\bibitem{} Leonard, P. J. T., \& Duncan, M. J. 1988, Astron. J., 96, 222 
\bibitem{} Leonard, P. J. T., \& Duncan, M. J. 1990, Astron. J., 99, 608 
\bibitem{} Lu, J. R., et al., 2005, Ap. J., 625, L51 
\bibitem{} Maeder, A. 1991, A \& A, 242, 93 
\bibitem{} Maillard, J. P., Paumard, T., Stolovy, S. R. \& Rigaut, F. 2004, A \& A, 423, 155
\bibitem{} Makino, J. \& Aarseth, S.J., 1992, PASJ 44, 141 
\bibitem{} Mas-Hesse, J. M., \& Kunth, D. 1991, A \& A Suppl. Ser., 88, 399 
\bibitem{} McCrady, N., Gilbert, A. M., \& Graham, J. R. 2003, Ap. J., 596, 240
\bibitem{} Meynet, G. 1995, A \& A, 298, 767
\bibitem{} Meynet, G., \& Maeder, A. 2003, A \& A, 404, 975
\bibitem{} Mikkola, S. \& Aarseth, S. J. 1993, Celest. Mech. Dyn. Astron., 57, 439
\bibitem{} Perna, R., \& Stella, L. 2004, Ap. J., 615, 222
\bibitem{} Pols, O. R., Cote, J., Waters, L. B. F. M., \& Heise, J. 1991, A \& A, 241, 419
\bibitem{} Portegies Zwart, S. F., \& McMillan, S. L. W. 2002, Ap. J., 576, 899
\bibitem{} Portegies Zwart, S. F., et al., 2004, Nature, 428, 724 
\bibitem{} Portegies Zwart, S. F., et al., 2006, Ap. J., 641, 319 
\bibitem{} Portegies Zwart, S. F., Makino, J., McMillan, S. L. W., \& Hut, P. 1999, A \& A, 348, 117 
\bibitem{} Poveda, A., Ruiz, J., \& Allen, C. 1967, Bol. Obs. Tonantzintla y Tacubaya, 4, 86
\bibitem{} Ptak, A., \& Colbert, E. 2004, Ap. J., 606, 291
\bibitem{} Schaerer, D., \& Vacca, W. D. 1998, Ap. J., 497, 618
\bibitem{} Soria, R. 2007, Astrophys. Space Sci., 311, 213   
\bibitem{} Suzuki, T.K., et al., 2007, Ap. J., 668, 435 
\bibitem{} Van Bever, J., Belkus, H., \& Vanbeveren, D., 2008, in preparation
\bibitem{} Van Bever, J., \& Vanbeveren, D. 1997, A \& A, 322, 116
\bibitem{} Van Bever, J., \& Vanbeveren, D. 2000, A \& A, 358, 462
\bibitem{} Van Bever, J., \& Vanbeveren, D. 2003, A \& A, 400, 63
\bibitem{} Vanbeveren, D., 1991, A \& A 252, 159
\bibitem{} Vanbeveren, D. 1995, A \& A, 294, 107
\bibitem{} Vanbeveren, D., \& De Loore, C. 1994, A \& A, 290, 129
\bibitem{} Vanbeveren, D., et al., 1998a, New Astron., 3, 443 
\bibitem{} Vanbeveren, D., Van Bever, J., \& Belkus, H. 2007, Ap. J., 662, L107
\bibitem{} Vanbeveren, D., Van Bever, J., \& De Donder, E. 1997, A \& A, 317, 487 
\bibitem{} Vanbeveren, D., De Loore, C., \& Van Rensbergen, W. 1998b, A \& A Rev., 9, 63
\bibitem{} Vanbeveren, D., Van Rensbergen, W., \& De Loore, C. 1998c, The Brightest Binaries, Boston: Kluwer
Academic 

\end{thebibliography}
\end{document}